\documentclass[runningheads,a4paper]{llncs}

\usepackage{amssymb}
\usepackage{graphicx}

\usepackage{url}
\urldef{\mailsa}\path|{alfred.hofmann, ursula.barth, ingrid.haas, frank.holzwarth,|
\urldef{\mailsb}\path|anna.kramer, leonie.kunz, christine.reiss, nicole.sator,|
\urldef{\mailsc}\path|erika.siebert-cole, peter.strasser, lncs}@springer.com|    
\newcommand{\keywords}[1]{\par\addvspace\baselineskip
\noindent\keywordname\enspace\ignorespaces#1}

\usepackage{graphics} 
\usepackage{epsfig} 

\begin{document}

\title{Algorithms for Computing Topological Invariants in 2D and 3D Digital Spaces}

\author{Li Chen 
}

\institute{ University of the District of Columbia\\
            lchen@udc.edu}

\maketitle
\thispagestyle{empty}

\begin{abstract}

Based on previous results of digital topology, this paper focuses on
algorithms of  topological invariants of objects in 2D and 3D Digital Spaces.
We specifically interest in solving hole counting of 2D objects and genus
of closed surface in 3D. We first prove a new formula for hole 
counting in 2D. The number of
of holes is $h=1 + (|C_4|-|C_2|)/4$ where $C_4$ and $C_2$ are
  sets of inward and outward corner points, respectively. 

This paper mainly deals with algorithm design and implementation of 
practical computation of topological invariants in digital space. 
The algorithms relating to data structures , and pathological case 
detection and original data modification are main issues.

This paper designed fast algorithms for topological invariants such as 
connected components, 
 hole counting in 2D and  boundary surface genus for 3D.
 For 2D images, we designed a linear time algorithm to solve hole counting
problem. In 3D, we designed also $O(n)$ time algorithm to get genus of 
the closed surface. These two algorithms are both in $O(\log n)$ space 
complexity.


\end{abstract}


\section{Introduction}
Topological properties for objects in 2D and 3D space is an important task
 in image processing. An interesting problem called hole counting 
 is to count 
the number of holes in an 2D image.  
On the other hand, 3D computer graphics and computational 
geometry have usually used triangulation to represent a
3D object. It uses the marching-cube algorithm to transfer a
digital object into the representation of simplicial complexes. It requires 
very large amount of space in computer memory.    

In recent years, 
the developments in medical imaging and 3D digital camera systems 
raise the problem of the direct 
treatment of digital 3D objects due to the speed concerns. In theory,
digital topology will provide the method for such fast calculation.  

This paper will design fast algorithms for those calculation based on 
digital topology. 

This paper provides a complete process that deals with simulated and real 
data in order to obtain
the topological invariants for 2D and 3D images. The algorithms are: 
(1) 2D hole counting, and (2) 3D boundary surface genus calculation.   

One of the most difficult parts in real image processing is to deal with
some noises or  pathological cases. This paper also gives detailed
procedures for detecting those cases. And we will provide the reasons to
modify the original image into the image where the mathematical formula 
could apply to.



\section{Some Concepts of Digital Space }

Digital topology was developed for image analysis. We 
especially interested in obtaining the topological
properties for 2D and 3D images, e.g., topological invariants of images.
In practice, hole counting in 2D and genus in 3D spaces are most 
popular problems for real world problems. Now, we first review some
concepts. 
 
\subsection{Basic Concept of Digital Space}
  
A digital space  is a discrete space in which each
point can be defined as an integer vector.  
Two-dimensional digital space $\Sigma_{2}$ first. A point $P (x, y)$ in 
$\Sigma_{2}$ has two
horizontal $(x, y\pm 1)$ and two vertical neighbors $(x\pm 1, y)$. These four
neighbors are called directly adjacent points of $p$ . $p$ has also four
diagonal neighbors: $(x\pm 1, y\pm 1)$. These eight (horizontal, vertical and
diagonal) neighbors are called general (or indirect)
adjacent
points of $p$ .

        Let $\Sigma_{m}$ be m-dimensional digital space. Two points
$p=(x_1, x_2,...,x_m)$ and $q=(y_1,y_2,...,y_m)$ in $\Sigma_{m}$ are
directly adjacent points, or we say that $p$ and $q$ are direct
neighbor if

\centerline {$ d_{D}(p,q)=\sum_{i=1}^m |x_i-y_i|=1 .$}

\noindent $p$ and $q$ are indirectly adjacent points if

\centerline {$ d_{I}(p,q)=\max_{1\le i\le m} |x_i-y_i|=1. $}

\noindent For instance, in 2D, $d_{D}$ means 4-adjacency, and 
$d_{I}$ is 8-adjacency. In 3D, $d_{D}$ means 6-adjacency, and 
$d_{I}$ is 26-adjacency.  

\noindent (Note: ``Indirectly adjacent points'' include all
directly adjacent points here. It may be the reason that we should
change the word of ``indirectly'' to ``generally.'')

        In a three-dimensional space $\Sigma_{3}$, a point has six directly
adjacent points and 26 indirectly adjacent points. Therefore, two directly
adjacent points in $\Sigma_{3}$ are also called 6-connected, while two
indirectly adjacent points are also called
26-connected.
In this note, we mainly consider the direct adjacency. If we omit the word ``direct,'' ``adjacency'' means the direct adjacency.

        A point in $\Sigma_{m}$ is called a point-cell or
0-cell.
A pair of points
$\{p,q\}$ in $\Sigma_{m}$ is called a line-cell\index{Line-cell} or
1-cell\index{1-cell}, if $p$ and $q$ are adjacent
points. A surface-cell\index{surface-cell} is a set of 4 points which form a unit square
parallel to coordinate planes. A 3-dimensional-cell (or 3-cell)\index{3-cell}
is a unit cube which includes 8 points. By the same reasoning,
we may define a $k$-cell.  Fig. 1(a)(b)(c)(d) show a point-cell, line-cell, a
surface-cell and a 3-cell, respectively.

        Now let us consider to the concepts of adjacency and connectedness of
(unit) cells. Two points $p$ and $q$ (point-cells, or 0-cells) areif there exists a simple path $p_0,p_1,...,p_n$, where $p_0=p$ and
$p_n=q$, and $p_i$ and $p_{i+1}$ are adjacent for $i=1,...,n-1 $.

       Two cells are point-adjacent\index{Point adjacent} if they share a point. For example,
line-cells $C1$ and $C2$ are point-adjacent in Fig. 1 (e), and
surface-cells $s1$ and $s2$ are point-adjacent in Fig. 1(f).
Two surface-cells are line-adjacent\index{Line adjacent} if they share a line-cell. For
example, surface-cells $s1$ and $s3$ in Fig. 1(g) are line-adjacent.

       Two line-cells are point-connected\index{Point connected} if they are two end elements
of a line-cells path in which each pair of adjacent line-cells
is point-adjacent. For example, line-cells $C1$ and $C3$ in Fig. 1(e) 
are point-connected. Two surface-cells are
 line-connected\index{Line connected} if they are two end elements of a surface-cells path
 in which each pair adjacent surface-cells
are point-adjacent.
For example, $s1$ and $s2$ in Fig. 1(f) are line-connected.

Two $k$-cells are $k'$-dimensional adjacent ($k'$-adjacent),
$k>k'\ge 0$, if they share a $k'$-dimensional cell.
A (simple) $k$-cells path
with $k'$-adjacency is a sequence of  $k$-cells $v_0,v_1,...,v_n$,
 where $v_i$ and $v_{i+1}$
 are $k'$-adjacent and  $v_0,v_1,...,v_n$
are different elements. Two $k$-cells are called $k'$-dimensional
connected if
they are two end elements of a (simple) $k$-cells path
with $k'$-adjacency.

Assume that $S$ is a subset of $\Sigma_{m}$.
Let $\Gamma^{(0)}(S)$ be the set of all points in $S$,
and $\Gamma^{(1)}(S)$ be the line-cells set in $S$,...,
$\Gamma^{(k)}(S)$\index{$\Gamma^{(k)}(S)$} be the set of $k$-cells of $S$.
We say two elements $p$ and $q$ in  $\Gamma^{(k)}(S)$ are $k'$-adjacent
if $p\cap q\in \Gamma^{(k')}(S)$, $k' < k$.

   Let $p\in \Sigma_{3}$, a line-neighborhood of $p$ is a set containing $p$
and its two adjacent points. A surface-neighborhood of $p$ is
 a (sub-)surface where
$p$ is a inner point of the (sub-)surface.

$\Sigma_{m}$ represents a special graph $\Sigma_{m}=(V,E)$.
$V$ contains all integer
grid points in the $m$ dimensional Euclidean space \cite{MR,Che04}.
The edge set $E$ of $\Sigma_{m}$
is defined as $E = \{(a,b) | a, b \in V \& d(a,b)=1\}$ , where $d(a,b)$the distance between $a$ and $b$. In fact, $E$ contains all
pairs of adjacent points.
Because $a$ is an $m$-dimensional vector, $(a,b)\in E$
means that only
one component, the $i$-th component, is different in $a$ and $b$,
$|x_i - y_i|=1$, and the
rest of the components are the same where
 $a=(x_1,...,x_m)$ and  $b =(y_1,...,y_m)$. This is known as
the direct adjacency. One can define indirect adjacency as
$\max_{i} |x_i - y_i|=1 $.
 $\Sigma_{m}$  is usually called an $m$-dimensional digital space.basic discrete geometric element $n$-cells can be defined in such a
space, such as 0-cells (point-cells), 1-cells (line-cells), and 2-cells
(surface-cells).


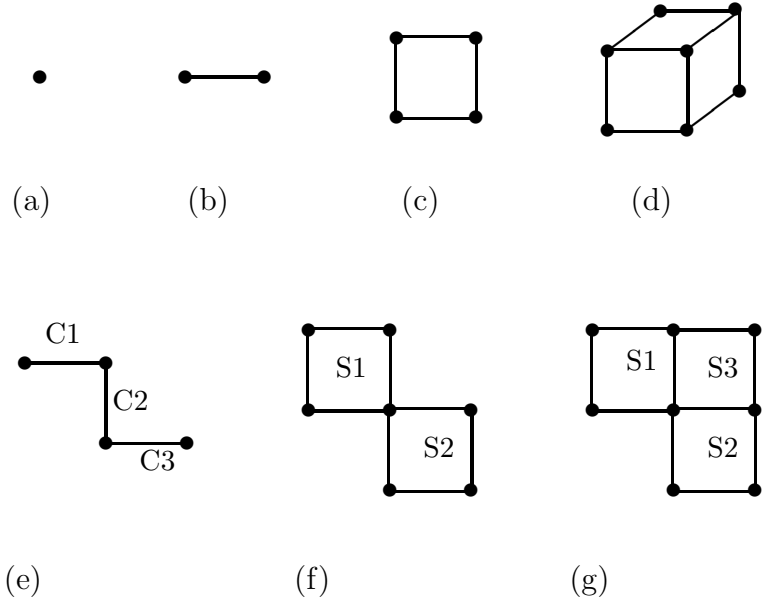
\begin{figure}[hbt]

\begin{center}
\setlength{\unitlength}{2500sp}%
\begingroup\makeatletter\ifx\SetFigFont\undefined%
\gdef\SetFigFont#1#2#3#4#5{%
  \reset@font\fontsize{#1}{#2pt}%
  \fontfamily{#3}\fontseries{#4}\fontshape{#5}%
  \selectfont}%
\fi\endgroup%
\begin{picture}(7448,5872)(2176,-6565)
\thicklines
\put(2531,-1446){\circle*{132}}
\put(3956,-1446){\circle*{132}}
\put(4732,-1446){\circle*{132}}
\put(8101,-1185){\circle*{132}}
\put(8101,-1970){\circle*{132}}
\put(8878,-1970){\circle*{132}}
\put(8878,-1185){\circle*{132}}
\put(9396,-1577){\circle*{132}}
\put(8620,-792){\circle*{132}}
\put(6028,-1054){\circle*{132}}
\put(6806,-1054){\circle*{132}}
\put(6806,-1839){\circle*{132}}
\put(6028,-1839){\circle*{132}}
\put(9351,-773){\circle*{132}}
\put(8101,-1970){\framebox(777,785){}}
\put(6028,-1839){\framebox(778,785){}}
\put(8878,-1185){\makebox(6.6667,10.0000){\SetFigFont{10}{12}{\rmdefault}{\mddefault}{\updefault}.}}
\put(8878,-1970){\makebox(6.6667,10.0000){\SetFigFont{10}{12}{\rmdefault}{\mddefault}{\updefault}.}}
\put(8101,-1185){\line( 4, 3){520.800}}
\put(8620,-792){\line( 1, 0){776}}
\put(9396,-792){\line( 0,-1){785}}
\put(9396,-1577){\line(-4,-3){520.160}}
\put(9396,-792){\line(-4,-3){520.160}}
\put(3956,-1446){\line( 1, 0){776}}
\put(7953,-3944){\circle*{132}}
\put(8748,-3944){\circle*{132}}
\put(8748,-4737){\circle*{132}}
\put(7953,-4737){\circle*{132}}
\put(9543,-4737){\circle*{132}}
\put(8748,-5532){\circle*{132}}
\put(9543,-5532){\circle*{132}}
\put(9543,-3944){\circle*{132}}
\put(7953,-4737){\framebox(795,793){}}
\put(8748,-5532){\framebox(795,795){}}
\put(8748,-3944){\line( 1, 0){729}}
\put(9477,-3944){\line( 1, 0){ 66}}
\put(9543,-3944){\line( 0,-1){793}}
\put(5169,-3944){\circle*{132}}
\put(5963,-3944){\circle*{132}}
\put(5963,-4737){\circle*{132}}
\put(5169,-4737){\circle*{132}}
\put(6759,-4737){\circle*{132}}
\put(5963,-5532){\circle*{132}}
\put(6759,-5532){\circle*{132}}
\put(5169,-4737){\framebox(794,793){}}
\put(5963,-5532){\framebox(796,795){}}
\put(2384,-4274){\circle*{132}}
\put(3179,-4274){\circle*{132}}
\put(3179,-5069){\circle*{132}}
\put(3974,-5069){\circle*{132}}
\put(2384,-4274){\line( 1, 0){795}}
\put(3179,-4274){\line( 0,-1){795}}
\put(3179,-5069){\line( 1, 0){795}}
\put(2583,-4075){\makebox(0,0)[lb]{\smash{\SetFigFont{11}{13.2}{\familydefault}{\mddefault}{\updefault}C1}}}
\put(3245,-4737){\makebox(0,0)[lb]{\smash{\SetFigFont{11}{13.2}{\familydefault}{\mddefault}{\updefault}C2}}}
\put(3511,-5334){\makebox(0,0)[lb]{\smash{\SetFigFont{11}{13.2}{\familydefault}{\mddefault}{\updefault}C3}}}
\put(5433,-4407){\makebox(0,0)[lb]{\smash{\SetFigFont{11}{13.2}{\familydefault}{\mddefault}{\updefault}S1}}}
\put(6295,-5202){\makebox(0,0)[lb]{\smash{\SetFigFont{11}{13.2}{\familydefault}{\mddefault}{\updefault}S2}}}
\put(9079,-5202){\makebox(0,0)[lb]{\smash{\SetFigFont{11}{13.2}{\familydefault}{\mddefault}{\updefault}S2}}}
\put(9079,-4407){\makebox(0,0)[lb]{\smash{\SetFigFont{11}{13.2}{\familydefault}{\mddefault}{\updefault}S3}}}
\put(8283,-4341){\makebox(0,0)[lb]{\smash{\SetFigFont{11}{13.2}{\familydefault}{\mddefault}{\updefault}S1}}}
\put(2251,-2761){\makebox(0,0)[lb]{\smash{\SetFigFont{12}{14.4}{\familydefault}{\mddefault}{\updefault}(a)  }}}
\put(2176,-6511){\makebox(0,0)[lb]{\smash{\SetFigFont{12}{14.4}{\familydefault}{\mddefault}{\updefault}(e)  }}}
\put(3976,-2761){\makebox(0,0)[lb]{\smash{\SetFigFont{12}{14.4}{\familydefault}{\mddefault}{\updefault}(b)  }}}
\put(6076,-2761){\makebox(0,0)[lb]{\smash{\SetFigFont{12}{14.4}{\familydefault}{\mddefault}{\updefault}(c)  }}}
\put(8326,-2761){\makebox(0,0)[lb]{\smash{\SetFigFont{12}{14.4}{\familydefault}{\mddefault}{\updefault}(d) }}}
\put(5026,-6511){\makebox(0,0)[lb]{\smash{\SetFigFont{12}{14.4}{\familydefault}{\mddefault}{\updefault}(f)  }}}
\put(7726,-6511){\makebox(0,0)[lb]{\smash{\SetFigFont{12}{14.4}{\familydefault}{\mddefault}{\updefault}(g)   }}}
\end{picture}
\end{center}
\caption{Examples of basic unit cells and their connections :
    (a) 0-cells, (b) 1-cells, (c) 2-cells, (d) 3-cells, (e) point-connected
    1-cells, (f) point-connected 2-cells, and (f) line-connected 2-cells.}
\end{figure}

\subsection{A Simple Lemma for 2D Digital Curve} 
We have proved some related theorem using Euler Characteristics and 
Gauss-Bonett Theorem.
The first is about simple closed digital curves. 

$C$ is a simple closed curve in direct (4-) adjacency where each element 
in $C$ is a point in  $\Sigma_{2}$ .   

We use $IN_{C}$ to represent the internal part of $C$. 
Since direct adjacency has the Jordan separation property,
$\Sigma_2-C$ will be disconnected. 

We also call a point $p$ on $C$ a $CP_{i}$ point if $p$ has
$i$ adjacent points in $IN_{C}\cup C$. In fact, $|CP_{1}|=0$ and
$|CP_{i}|=0$ if $i>4$ in $C$.

$CP_{2}$ contains outward corner points, $CP_{3}$ contains straight-line points, and
$CP_{4}$ contains inward corner points. 

For example, the following center point is
a outward corner point in array (Also see Fig. 2 ):
\[
\begin{array}{lll}
0 & 0 & 0 \\
0 & 1 & 1 \\
0 & 1 & x 
\end{array} \]

\noindent But in next array, the center point is an inward corner point:  

\[ \begin{array}{lll}
0 & 1 & x \\
1 & 1 & x \\
x & x & x 
\end{array} \]

\begin{figure}[h]
   \begin{center}
   \epsfxsize=3in 
   \epsfbox{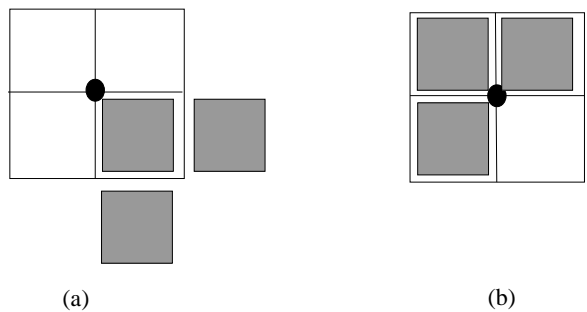}
   \end{center}
\caption{The outward corner point and inward  corner point}
\end{figure}

In ~\cite{Che04} (on page 20) , we used the Euler theorem to show a
result for a simple closed curve $C$, 
 
\begin{lemma} 
\begin{equation}
      CP_2 = CP_4 + 4.
\end{equation}
\end{lemma}

\subsection{Genus of Digital Surfaces in 3D}
In this section, we review some existing work related to this paper including the genus
of closed digital surfaces, homology groups of manifolds in 3D digital space, and
a theoretical linear algorithm of finding Homology Groups in 3D~\cite {CR08}.

Any continuous 3D object can be viewed as a collection of 3D voxels in digital or
cubical space.
Unless the sampling method is changed, any practical method of genus calculation must adapt to this fact. 
Medical imaging such as CT and MRI are such examples.     

Cubical space with direct adjacency, or (6,26)-connectivity in digital space\cite{Che04}, has the simplest 
topology in 3D digital spaces. It is also sufficient for the topological 
property extraction of 
digital objects in 3D. Two points are said to be adjacent in 
(6,26)-connectivity space if the 
Euclidean distance of these two points is 1, called direct adjacency.
 
Let $M$ be a closed (orientable) digital surface in the 3D grid space in direct adjacency. 
We know that there are exactly 6-types of digital surface
points (See Fig. 3). This was discovered by Chen and Zhang in~\cite{CZ93}. 
Relation to different 
definitions of digital surfaces  can be found in 
~\cite{CCZ99,Che04}.

\begin{figure}[h]
   \begin{center}

   \epsfxsize=2in 
   \epsfbox{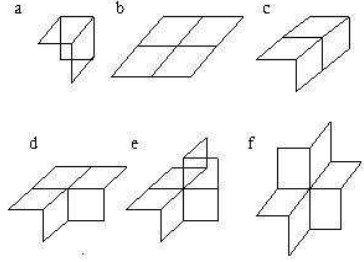}

   \end{center}
\caption{ Six types of digital surfaces points in 3D}
\end{figure}

Assume that $M_i$ ($M_3$, $M_4$, $M_5$, $M_6$) is the set of 
digital points with $i$ 
neighbors.   We have the following result for a simply connected 
$M$ ~\cite{CZ93}\cite{Che04}:

\begin{equation}
          |M_3| =8 + |M_5| + 2 |M_6| .      
\end{equation}
 
\noindent $M_4$ and $M_6$ has two different types, respectively.
Gauss-Bonnet theorem states that if $M$ is a closed manifold, then

\begin{equation}
          \int_{M} K_{G} d A = 2 \pi \chi(M)    
\end{equation}

\noindent where $d A$ is an element of area and $K_{G}$ is the Gaussian curvature. 

Its discrete form is

\begin{equation}
          \Sigma_{\{p \mbox{ is a point in } M\}} K(p) = 2 \pi \cdot (2- 2 g)  
\end{equation}


\noindent where $g$ is the genus of $M$.

Assume that $K_i$ is the curvature of elements in $M_i$, $i=$ 3,4,5,6. We have

\begin{lemma}\label{l21}
   (a) $K_3  = \pi/2$,
 (b) $K_4= 0$,  for both types of digital surface points,
  (c) $K_5  = - \pi /2$, and
   (d) $K_6  = - \pi$,  for both types of digital surface points.
\end{lemma}

\noindent We obtained (see \cite{CR08}),

\begin{equation}
           g = 1+ (|M_5|+2 \cdot |M_6| -|M_3|)/8. 
\end{equation} 

The three simple examples show that the above formula  is correct \cite{CR08}. 
See Fig. 4.  Different surface points can also be used
to form a feature vector for 3D surfaces. We have used 
it in face modeling~\cite{CS}.

\begin{figure}[h]
   \begin{center}

   \epsfxsize=2in 
   \epsfbox{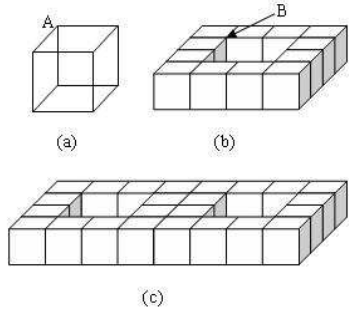}

   \end{center}
\caption{ Simple examples of closed surfaces with $g=0,1,2$}
\end{figure}

\begin{figure}[h]
   \begin{center}

   \epsfxsize=2.3in 
   \epsfbox{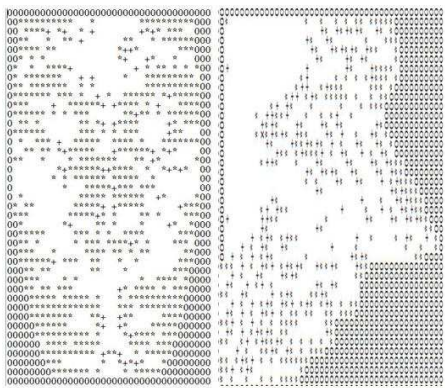}

   \end{center}
\caption{Different surface points on human faces.}
\end{figure}

For a $k$-manifold,
Homology group $H_{i}$, $i=0,...,k$ provides the information for the number of holes in each $i$-skeleton of the
manifold. When the genus of a closed surface is obtained, we can then
calculate the homology groups corresponding to its 3-dimensional
manifold in 3D. 

The following result follows from standard results in algebraic topology  \cite{Hat}.
It also appears in \cite{Day98}.  Let $b_i = \mbox{rank} H_i(M, Z)$ be the $i$th Betti number of $M$. 
The Euler characteristic of $M$ 
is defined by

\[ \chi (M) = \sum_{i\geq 0} (-1)^i b_i \]

If $M$ is a 3-dimensional manifold, $H_i(M)=0$ for all $i>3$ essentially because 
there are no $i$-dimensional holes.
Therefore, $\chi (M)  = b_o - b_1 + b_2 - b_3$.
 Furthermore, if $M$ is in
$R^3$, it must have nonempty boundary. This implies that $b_3 = 0$.

\begin{theorem}\label{Jordan2}
Let $M$ be a compact connected 3-manifold in $S^3$. Then
\begin{enumerate}
\item [(a)]  $H_0(M)\cong Z$.
 \item [(b)] $H_1(M) \cong Z^{\frac{1}{2} b_1(\partial M)}$, i.e. $H_1(M)$ is torsion-free with rank being half of rank $H_1(\partial M)$.
\item[(c)] $H_2(M) \cong Z^{n-1}$ where  $n$ is the number of components of $\partial M$.
\item[(d)] $H_3(M)=0$ unless $M=S^3$.
\end{enumerate}
\end{theorem}

A proof of above theorem is shown in \cite{CR08}.





\section{Hole Counting Problem in 2D}

In an image, hole counting 
is to count the number of holes in an 2D image. 
It was studied by some researchers before~\cite{Qia,LYS}. 
In this paper, we will get the simplest method to solve the problem.

A line or curve in real world always have a thickness no matter how
thin it is. However,
a digital line could give human some wrong interpretation. The example
in Fig. 6  will show you how similar digital objects produce different answers.

\begin{figure}[h]
   \begin{center}

   \epsfxsize=4in 
   \epsfbox{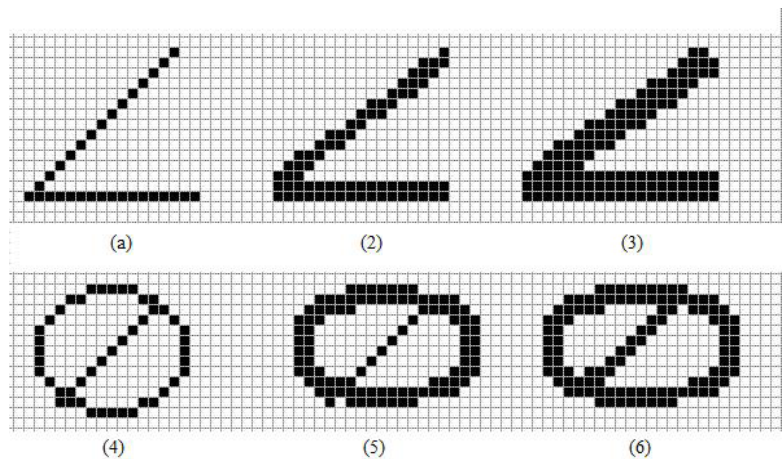}

  \end{center}
\caption{2D digital objects looks the same but topologically different:
         (1) an  angle with $thickness=1$ by MS Paint Software,
         (2) an angle with $thickness=2$,
         (3) an angle with $thickness=3$,
         (4) an ellipse and a line with $thickness=1$,
         (5) an ellipse and a line with different thickness,
         and (6) an ellipse and a line with $thickness=2$. 
 }
\end{figure}

Micro-software engineers for MS Paint probably never thought that there are 
some major difference except the width of a paint brash. The artistic
effect are not very different. However, the images in digital world makes
great deal of differentiality from Fig 6. (1) to (2) than the difference
from (2) to (3).  This is because that   Fig 6. (1) can be interpreted 
as a (square-)dotted line for one of its leg for direct adjacency. 
 A dotted line is a 
collection of several disconnected objects. This is much complicated
issue than a connected line. Fig. 6 (4) may be no hole, 
one hole, or two holes. In 8-adjacency, there is no hole. In 4-adjacency,
the points are not connected. The best way is to use 8-adjacency for
"1"s (foreground) and use 4-adjacency for "0"s (background). 
Only Fig. 6 (6) will give the
answer exactly the same as humans interpretation, i.e. two holes.  

If we use 8-adjacency for
"1"s and use 4-adjacency for "0"s, this type of adjacency is called
(8,4)-adjacency. It may cause another problem, for instance,
if we have two parallel "1" lines with a "0" line in between in 45 degree.
Each "0" point will be determined as a separated component. They are
not formed a connected "0" line. That is also against the human 
interpretation.  

In this paper, our method will assume that $C$ does not contain 
the following cases (if there is any, we will modify the
original image to remove them, we will discuss next):

\[\begin{array}{ll}
1 & 0 \\
0 & 1
\end{array} \]

\noindent and

\[\begin{array}{ll}
0 & 1 \\
1 & 0
\end{array}\]

\noindent These two cases are called  the pathological cases (See Fig.7)

\begin{figure}[h]
   \begin{center}
   \epsfxsize=2in 
   \epsfbox{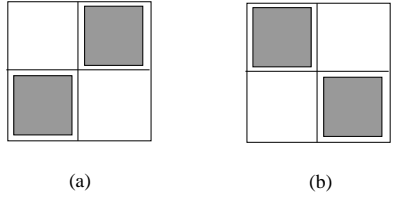}
   \end{center}
\caption{Two pathological cases  }
\end{figure}

It is obvious that, our paper does not solve all problems. It is too 
strict. The advantage of our method is to get a simple treatment.
In ~\cite{BNB}, there were detailed discussions about complex cases.

Our algorithm will fill or delete some points in the original image 
to make the pathological cases go away. 
We also want to remove single point whatever it is black or white. it 
will be treated as noise. 

Topological invariants should maintain Jordan property. We shall only
allow the direct adjacency in order to dealing with Topological invariants,
at least in most of cases.

The number of holes in a connected component in 2D images is a basic invariant. 
In this note, a simple formula was proven using our previous results in digital 
topology~\cite{Che04,CR}. The new is: $h =1+ (|C_4|-|C_2|)/4$ , where h is the number 
of holes, and
$C_i$ indicate the set of corner points having $i$ direct adjacent 
points in the component.


\subsection{The Simple Formula for the Number of Holes in $S$}

An image segmentation method can extract a connected component.  
A connected component $S$ in a 2D digital image is often used to represent
a real object. The identification of the object can be first done by determining
how many holes in the component. For example, letter ``A'' has one hole
and ``B'' has two holes. In other words, if $S$ has $h$ holes, then the complement of $S$ 
has $h+1$ connected components (if S does not reach the boundary of the image).

\begin{theorem}
Let $S \subset \Sigma_2$ be a connected component and its boundary $B$ is a
collection of simple closed curves  without pathological cases.  
Then, the number of holes in $S$ is
\begin{equation}
   h= 1+(C_4-C_2)/4
\end{equation}
\noindent  $C_4, C_2 \subset B$.
\end{theorem}

In this section, we will give two methods to prove
our result.  First, use the 3D formula to get the 
theorem for holes. Second, we use 2D formula to directly
prove the theorem.\\

{\bf \it The First Proof: } 

Let $S\subset \Sigma_2$ be a connected component and its boundary do not have the pathological 
cases.  (We actually can detect those cases in linear time.)

We can embed $S$ into $\Sigma_3$ to make a double $S$ in $\Sigma_3$. 
At $z=1$ plane, we have a $S$, denoted $S_1$, and we also have the exact same
$S$ at $z=2$ plane, denoted $S_2$. 

Without loss generality, $S_1\cup S_2$ is a solid object. (We here omit some
technical details for the strict definition of digital surfaces.) It's boundary
is closed digital surfaces with genus $g=h$. We know   
\[g = 1+ (|M_5|+2 \cdot |M_6| -|M_3|)/8\] 

There will be no points in $M_6$. 

For each point $x$ in $C_2$ in $C\subset S$ ($C$ is the boundary of $S$), 
we will get
two points in $M_3$ in $S_1\cup S_2$. In the same way, if a point $y$ is
inward in $C_4\in C$, we will get
two points in $M_5$ in $S_1\cup S_2$. There is no point in $M_6$, i,e.,
$|M_6|=0$. 
So $2|C_2|=|M_3|$, and $2|C_4|=|M_5|$. We have

\[h=g = 1+ (|M_5|+2 \cdot |M_6| -|M_3|)/8= 1+ (2|C_4|-2|C_2|)/8 \]

\noindent Thus,

\[h= 1+ (|C_4|-|C_2|)/4 \]
$\diamond$

We can also prove this theorem using Lemma 1 for digital curves:
$CP_2=CP_4+4$ for a simple closed curve. \\ 

{\bf \it The Second Proof: }
  
This can also be proved by the lemma in above section. 
\[CP_2=CP_4+4\]  

A 2D connected component $S$ with $h$ holes that contains 
$h+1$ simple closed curves in the boundary of $S$
Those curves do not cross each other. 

The $h$ curves corresponding to $h$ holes will be considered 
oppositely in terms of inward-outward. 

including one counts at inward and h is reversed
outward with inward. It will get there. 

Let $CP^{(0)}$ the outside curve of $S$ and  
$CP^{(i)}, i=1,\cdots,h$, is the curve for the $i$-th hole.  

Inward points to $S$ is the outward points to $C^{(i)}, i=1,\cdots,h$.
And vise versa.

\[CP_2^{(0)}=CP_4^{(0)}+4\]

\[CP_2^{(i)}=CP_4^{(i)}+4\]

The total outward points in the boundary of $S$ is

 \[ CP_{2}=CP_2^{(0)} + \sum_{i=1}^{h} CP_4^{(i)}\]

\noindent The inward points in the boundary of $S$ is
  
 \[CP_{4}=CP_4^{(0)} + \sum_{i=1}^{h} CP_2^{(i)}\]

\noindent Thus,

 \[CP_{4}-CP_{2}=CP_4^{(0)} + \sum_{i=1}^{h} CP_2^{(i)} 
                -CP_2^{(0)} - \sum_{i=1}^{h} CP_4^{(i)}\]

we have 
 $CP_{4}-CP_{2}=-4  + \sum_{i=1}^{h} 4  = -4 +4h$
 
Therefore, 

 \[h = 1 + (CP_{4}-CP_{2})/4\] 
 $\diamond$

We can see that this formula is so simple to get the holes 
(genus) for a 2D object
without any little sophistic algorithm, just count if the point is
a corner point, inward or outward. 

We could not get the similar simple formula in triangulated 
representation of the 2D object. This is the beauty of digital 
geometry and topology!

{\bf Example 1}
To test if this formula is correct, we select the following examples

\begin{eqnarray}
    \left (\begin{array}{llllllll}

0 & 0 & 0 & 0 & 0 & 0 & 0 & 0\\
0 & 0 & 1 & 1 & 1 & 1 & 0 & 0   \\
0 & 1 & 1 & 1 & 1 & 1 & 0 & 0 \\
0 & 1 & 1 & 1 & 0 & 0 & 0 & 0\\
0 & 0 & 1 & 1 & 0 & 0 & 0 & 0   \\
0 & 0 & 1 & 1 & 1 & 0 & 0 & 0\\
0 & 0 & 1 & 1 & 1 & 0 & 0 & 0   \\
0 & 0 & 0 & 0 & 0 & 0 & 0 & 0
     
              \end{array}
              \right )
\end{eqnarray}

In order to see clearly, we use ``2'' to represent points in $CP_2$ and
use ``4'' to represent points in $CP_4$.

\begin{eqnarray}
    \left (\begin{array}{llllllll}

0 & 0 & 0 & 0 & 0 & 0 & 0 & 0\\
0 & 0 & 2 & 1 & 1 & 2 & 0 & 0   \\
0 & 2 & 4 & 4 & 1 & 2 & 0 & 0 \\
0 & 2 & 4 & 1 & 0 & 0 & 0 & 0\\
0 & 0 & 1 & 1 & 0 & 0 & 0 & 0   \\
0 & 0 & 1 & 4 & 2 & 0 & 0 & 0\\
0 & 0 & 2 & 1 & 2 & 0 & 0 & 0   \\
0 & 0 & 0 & 0 & 0 & 0 & 0 & 0
     
              \end{array}
              \right )
\end{eqnarray}

In this example $|CP_2| = 8$ and $|CP_4| = 4$.  
 $h = 1 + (CP_{4}-CP_{2})/4 = 1 + (4-8)/4 =0$.

Another example is the following

\begin{eqnarray}
    \left (\begin{array}{llllllll}

0 & 0 & 0 & 0 & 0 & 0 & 0 & 0\\
0 & 0 & 1 & 1 & 1 & 1 & 1 & 1   \\
0 & 1 & 1 & 1 & 1 & 1 & 1 & 1 \\
0 & 1 & 1 & 1 & 0 & 0 & 1 & 1\\
0 & 1 & 1 & 1 & 0 & 0 & 1 & 1   \\
0 & 0 & 1 & 1 & 1 & 1 & 1 & 1\\
0 & 0 & 1 & 1 & 1 & 1 & 1 & 1   \\
0 & 0 & 0 & 0 & 0 & 0 & 0 & 0
     
              \end{array}
              \right )
\end{eqnarray}

In the second example $|CP_2| = 6$ and $|CP_4| = 6$.  
 $h = 1 + (CP_{4}-CP_{2})/4 = 1 + (6-6)/4 =1$.
 
\noindent When add a hole, we will add 4 more $CP_4$ points. That is the
reason why this formula is correct.

\subsection{Algorithms for Hole Counting}

The key of the algorithm is to delete all 
the pathological cases. There two types of such cases. Sometimes
it is hard to decide if we need to add a point or delete a point to a pathological case to be removed. 

The first method will based on the original grade space.
Sometimes, delete a pixel for removing
pathological case may add another pathological case. In such
a case, we will add a pixel to make the original pathological
case go away.    and vice versa.

In other case,
if add and delete will not reduce the pathological case, use delete.
since delete will eventually complete the job. It must be the
case shown in Fig. 8.   

\begin{figure}[h]
   \begin{center}

   \epsfxsize=2in 
   \epsfbox{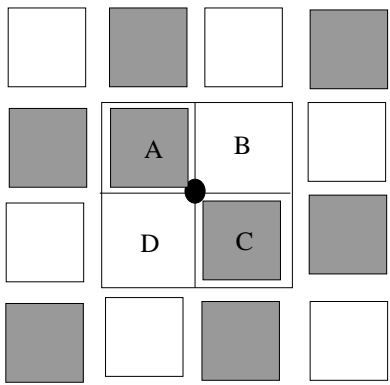}

   \end{center}
\caption{ Any deleting or adding a pixel will add a new pathological case }
\end{figure}

The second method is made to apply  to the case
that so many instances in the image as shown in Fig. 8.  
We need to consider a refinement method e.g. half grade space to avoid
deletion and addition of a cell that will cause other pathological 
case. This method will increase the space need. We use half size 
cells to fill the space. Or we just delete a half cell to remove 
the pathological case.\\

{\bf Algorithm 3.1}  Algorithm for calculating number of holes in 2D images.

\begin{description}

\item [Step 1.] Get connected components using direct-adjacency (4-adjacency). 

\item [Step 2.] Extract every component, do following. Fill all single "0" pixel or (unit-square or 2-cell)
            to be "1"  if all its indirect neighbors (8-adjacency) are "1".
            Delete all single "1" pixel 
            if all its indirect neighbors (8-adjacency) are "0". 
             
\item [Step 3.] Find all  pathological cases, delete or add a pixel 
            if this action does not create a new pathological case. 
            If either action will create a new pathological case, use
            delete.  Repeat this step until all pathological cases are 
            removed. 
                
\item [Step 4.] Count all inward edge points $CP_4$ and outward points
              $CP_{2}$ .

\item [Step 5.] $h = 1 + (CP_{4}-CP_{2})/4$
              
\end{description}

Note that deleting a pixel means to   
change the single "0" pixel (unit-square or 2-cell) to "1" pixel.
is for noise image. It is not very necessary for doing this step.
The above calculation is for each component to get holes. 
Algorithm 3.1 is an $O(\log n)$ space algorithm without Step 1
since it is just count the number of point types. If we assume
that the number of  pathological cases are constant, this algorithm
will be linear in time.

\section {Algorithms for Genus on 3D Digital Surfaces }

Basically, the topological properties of an object in 3D contains connected components,
genus of its boundary surfaces, and other homologic and homotopic properties~\cite {Kaz03}. 
In 3D, the problem of obtaining fundamental groups is decidable but no practical 
algorithm has yet been found.
Therefore, homology groups have played the most significant role~\cite{Day98} ~\cite{Kaz04} . 
 
Theoretical results show that there exist linear time algorithms for calculating genus and homology groups 
for 3D Objects in 3D space~\cite{Day98}. However, the implementation of these algorithms is not simple due to the complexity 
of real data samplings. Most of the algorithms require the triangulation of the input data since it is 
collected discretely. However, for most medical images, the data was sampled consecutively, meaning that
every voxel in 3D space will contain data. In such cases, researchers use the marching-cubes algorithm obtain the triangulation since it is 
a linear time algorithm ~\cite{LC87}. However the spatial 
requirements for such a treatment will be at least doubled by
adding the surface-elements (sometimes called faces). 

The theoretical work of calculating genus based on simple
decomposition will turn into two different
procedures: (1) finding the boundary of a 3D object and then using polygon mapping, also called polygonal schema, 
(2) cell complex reductions
where a special data structure will be needed.

In this paper, we look at a set of points in 3D digital space, and
our purpose is to find homology groups of the data set. 
The direct algorithm without utilizing triangulation was proposed by Chen and Rong in 2008~\cite {CR08}
However, this algorithm 
is based on the strict definition of digital surfaces. Many real 3D sets may not satisfy the definition. 
In other words, a set of connected points may not be able to be put 
into such a process without considerable associated theoretical
and practical processes. 

In \cite {CR08}, we discuss the geometric and algebraic properties of manifolds 
in 3D digital spaces and the optimal algorithms for calculating these properties. 
We consider {\em digital manifolds} as defined in ~\cite {Che04}. 
More information related to digital geometry and topology 
can be found in ~\cite{KR} and ~\cite{KoR}.
We presented a theoretical optimal algorithm with
time complexity $O(n)$ to compute the genus and homology groups in 3D digital space, 
where $n$ is the size of the input data ~\cite {CR08}.  
  
The key in the algorithm in ~\cite{CR08} is to find the genus of the closed digital surfaces that
is the boundary of the 3D object.
However, the new algorithm is based on the strict definition of closed digital surfaces in \cite{Che04},
which means that there are many cases of real sampling of 3D objects that do not satisfy the definition
of digital surfaces. In this paper, we will also deal with extreme situations. We have designed an adding and deleting method to make the 3D object into manifolds. 

first describe a theoretical procedure in this section for 
3D genus for 3D objects. The implementation of the algorithm must consider
all possible cases in practical data collection. We first need to find the boundary and then decide if
the boundary is a 2D manifold. If the boundary data connecting voxel data
sets are not purely defined digital surfaces, we will have three options: (1) we need to modify the data 
to meet the requirement before genus calculation, (2) if the change of the original data set is too great, we
may need to stop the modification instead of outputting a result for reference, and (3) we make some limited changes,
and then produce a result.     

The difference between the theoretical results and practical data processing is that we may not always get the 
input data we expected. In our case, the boundary of a solid object should be
treated as a surface. However, practically, this might not always be the case. 
Some researchers also consider making real data sets ``well''-organized.  Siqueira {\it et al}
considered making a 26-connected data set well-composed~\cite{SLG05} 
\cite{SLTGG08}~\cite{BB08}. This means that two voxels will be connected
by a sequence of voxels where each pair of two adjacent cubes share a 2D-cell (face-unit).    
The concept of well-composed is mathematically equivalent to 
6-connected. An algorithm described in ~\cite{SLG05} 
\cite{SLTGG08} may generate new "none" well-composed cases, which are not good selections 
 for genus calculation.

Our new algorithm and implementation will perform: (1) pathological cases detection and deletion, 
(2) raster space to point space (dual space) transformation, (3) the linear time algorithm for 
boundary point classification, and (4) genus calculation.    

Some detailed considerations of recognition algorithms related to 3D manifolds can be 
found in ~\cite{BK08} where
Brimkov and Klette made extensive investigations in boundary tracking.  The discussions 
of 3D objects in raster space can be found in ~\cite{Lat}.

\subsection{A Theoretical Procedure for Genus in 3D}

Based on the results we presented in the above subsections, we
now describe a linear algorithm for computing the homology group of 3D objects
in 3D digital space ~\cite{CR08}.

Assuming we only have a set of points in 3D. We can digitize this set into 3D digital spaces. 
There are two ways of doing so: (1) by treating each point as a cube-unit that is called the 
raster space, 
(2) by treating each point as a grid point, which is also called the point space.
These two are dual spaces.
Using the algorithm described in ~\cite{Che04}, we can determine whether the digitized set forms a 
3D manifold in 3D space in direct adjacency for connectivity. The algorithm is in linear time. \\

\noindent {\bf Algorithm 4.1} Let us assume that we have a connected $M$ that 
is a 3D digital manifold in 3D.

\begin{description}

 \item [Step 1.] Track the boundary of $M$, $\partial M$, which is a union of several closed surfaces. 
This algorithm only needs to scan though all the points in $M$ to see if 
the point is linked to  
a point outside of $M$. That point will be on boundary. 

\item [Step 2.]  Calculate the genus of each closed surface in $\partial M$ using the method 
described in Section 2. We just need to count the number of neighbors on a surface.  
and put them in $M_i$, using the formula (5) to obtain $g$.

\item [Step 3.] Using the Theorem \ref{Jordan2}, we can get $H_0$, $H_1$, $H_2$, and $H_3$. 
                $H_0$ is $Z$. For $H_1$, we need to get $b_1(\partial M)$ that is just 
            the summation of the genus in all connected components in  $\partial M$. (See \cite{Hat}
            and \cite{Day98}.) 
            $H_2$ is the number of components in $\partial M$. $H_3$ is trivial.
\end{description}

\begin{lemma}
           Algorithm 4.1 is a linear time algorithm. 
\end{lemma}

Therefore, we can use linear time algorithms to calculate $g$ and all homology
 groups for digital manifolds in 3D based on Lemma 2, Formula (5) and Lemma 3.

\begin{theorem}
           There is a linear time algorithm to calculate all homology
 groups for each type of manifold in 3D.  
\end{theorem}

\subsection{3D Input Data Sets}

This subsection will discuss the input data formats. We will focus on cubical data, 
for instance MRI and CT data.
In cubical data samples, we assume that the sampling is contiguous, where each sample point is 
normally followed by 
another sample point in its neighborhood. It is important to know this because a random sampling can cause 
the problem of uncertainty.  
In this case, we usually
cannot calculate the genus without making an assumption. For instance, we will not be able to know 
where a hole is. 
In order to get simplicial decomposition (usually triangulation), we usually need to use 
Voronoi or Delaunay decomposition
with boundary information. That means the boundary must be assumed. 

A new technology is called persistent homology analysis that tells us how to find the best estimation
 for the location 
of holes, usually by multiscaling (the upscaling and downscaling methods). However, this method is not a 
precise analysis ~\cite{Car}~\cite{ZC}.  
  
Even though, our method can be modified to be used in persistent analysis, this paper mainly deals with
the method of precise genus and homology group calculation. 

In summary, our assumption is that the digital object consists of cubical points (digital points, raster points). 
Each point is a cube, which is the smallest 3D object. The edge and point are defined with regards to the cube and
an object may contain
several connected components using a cube-linking path. Our purpose again is to calculate the 
topological properties
of the object, or of each component, essentially. 

\subsection{Searching connected components of a cubical data set}

Connected component search is an old task that can be done by using Tarjan's Breadth-first-search. 
Pavlidis was one of  
the first people to realize and use this algorithm in image processing. This problem is also known as the 
labeling problem. The complexity of the algorithm is $O(n)$ ~\cite{Pav}. 

The problem is what connectivity is based off of. In 3D, we usually have 6-, 18-, 26- connectivity. Since real data has
noise, we have to consider all of those connectivities. So we must use 26-connectivety to get the connected components.

Therefore, the connected component of the real processing is not a strictly 6-connected component. The topological 
theorem generated previously in ~\cite{CR08} is no longer suitable. So we need to transform 
a 26-connected component into a 6-connected component. This should be done by a meaningful adding or deleting process since
optimization on the minimum number of changes could be an NP-hard problem. 

{\bf Problem of minimum modifications:} Given a set of points in 3D digital space, if this set is not a manifold, assume that the points are connected in 
a connectivity defined using adding or deleting processes to make the set
a 3D manifold. The question becomes: is there a polynomial algorithm that makes  the solution have minimum modifications where adding or deleting 
 a data point will be counted as one modification?

A similar problem was considered in ~\cite{SLG05} in which a decision problem of adding was proposed. 

This problem can be extended to a general $k$-manifold in $n$-D space. Even though we have the 6-connected 
component, there may still be cases that 
contain the pathological situation, which needs special treatment. We will discuss this issue in 
the next subsection.

\subsection{Pathological Cases Detection and Deletion}

In this paper, we only deal with the Jordan manifolds, 
meaning that a closed $(n-1)$-manifold will separate the $n$-manifold into two or more components. 
For such a case, only direct adjacency will be allowed since indirect adjacency will not generate
 Jordan cases. 

That is to say, if the set contains indirect adjacent voxels, 
we need to design an algorithm to detect the situation and delete some voxels in order to preserve  the homology groups.

It is known that there are only two such cases in cubical or digital space \cite{Che04}: two 
voxels (3-cells) share 
a 0-cell or a 1-cell. Therefore, we want to modify the voxel set to only contain voxels where two of these cases 
do not appear. Two voxels share exactly a 2-cell, or there is a local path (in the neighborhood) voxels where two adjacent
voxels share a 2-cell ~\cite{Che04}. A special case was found in \cite{SLTGG08} that is the complement case of
the case in which two voxels share a 0-cell (see Fig 9. (a)). 
The case may create a tunnel or could be filled. We will simplify it by 
adding a voxel in a $2\times 2\times 2$ cube. 
Such a case in point space is similar to the case (a) in Fig 9 since 
boundaries of these cases are the same.  
 
The problem is that many real data sets do not satisfy the above restrictions (also called well composed
image). The detection is easy but deleting
certain points (the minimum points deletion) to preserve the homology is a bigger issue.

The following rules (observations) are reasonable: In a neighborhood $N_{27}(p)$ that contains 8 cubes and
27 grid points,

 a) if a voxel only shares a 0-cell with a voxel. This voxel can be deleted.  

 b) if a voxel only shares a 1-cell with a voxel. This voxel can be deleted.  

 c) if a boundary voxel $v$ shares a 0,1-cell with a voxel, assume $v$ also shares a 2-cell with a voxel $u$, 
    $u$ must share a 0,1-cell with a voxel that is not in the object $M$.  $u$ is on the boundary. 
    Deleting $v$ will not change the topological properties.
    
 d) if in a $2\times 2\times 2$ cube, there are 6 boundary voxels and 
   its complement (two zero-valued voxels) 
    is the case (a) in Fig. 9. Add a voxel to this 
	 $2\times 2\times 2$ cube such that the new voxel
    shares as many 2-cells in the set as possible. 
	 This means that we want the adding voxel to be inside of
    the object as much as we can.

\begin{figure}[h]
   \begin{center}

   \epsfxsize=3in 
   \epsfbox{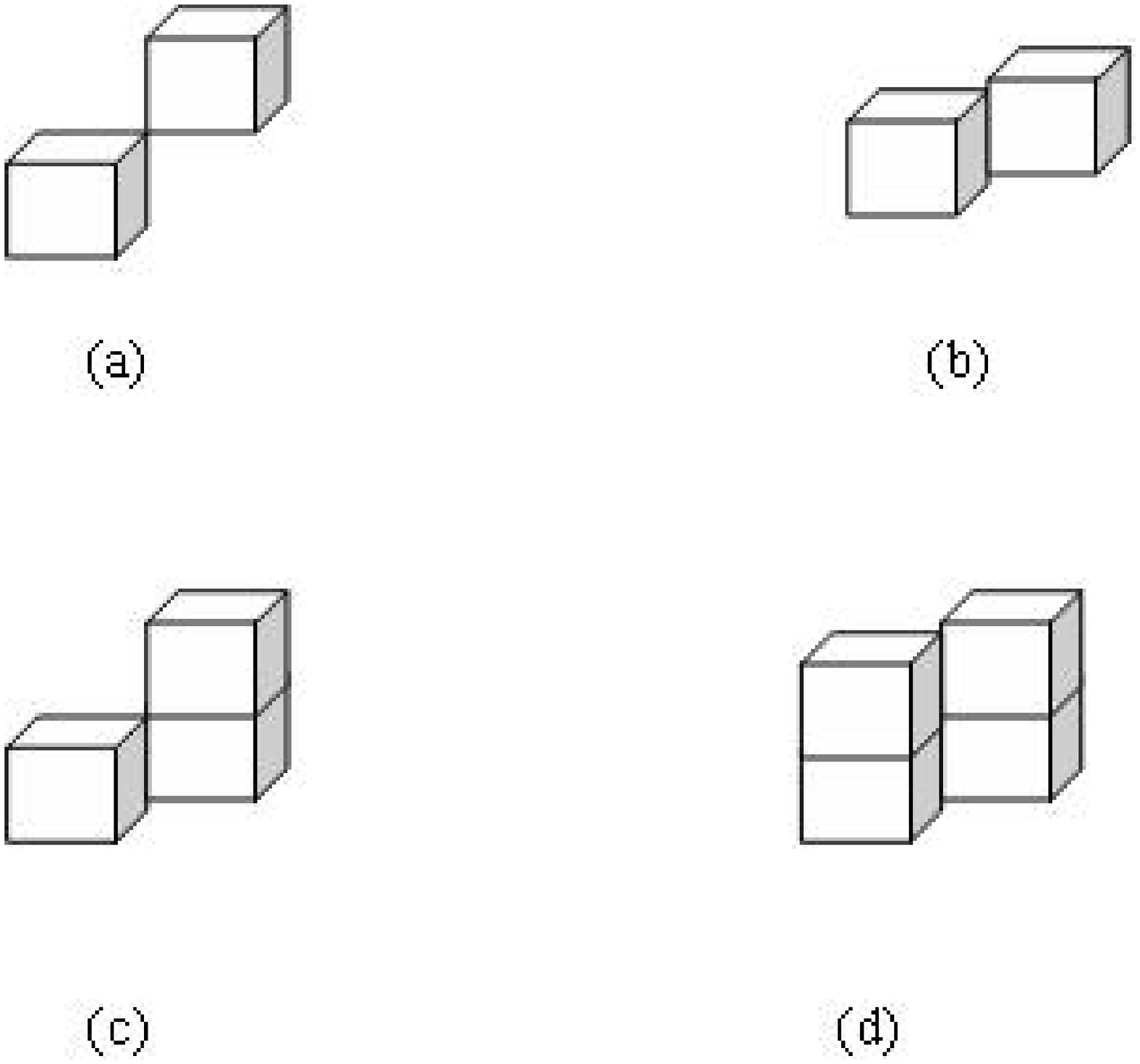 }

   \end{center}
\caption{ Pathological Cases in 3D}
\end{figure}

In this paper, we implement or modify the above rules to fit the 
theoretical definition of the digital surfaces.  We also design an algorithm based on these 
rules to detect and delete some data points while preserving 
the topology. This is essential to calculating the genus correctly.  However, when the object becomes more
complex, pathological situations may still exist.

\begin{figure}[h]
   \begin{center}

   \epsfxsize=4in 
   \epsfbox{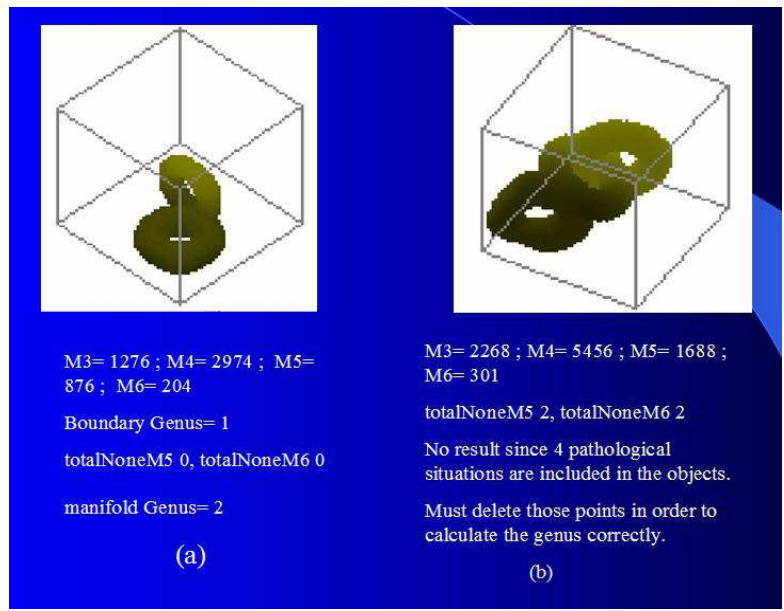}

   \end{center}
\caption{ Without Pathological Case Process}
\end{figure} 

The mathematical foundation of this above process that eliminates pathological cases is still under investigation. 

{\bf Mathematical foundation of Modifying a 3D object to be a 3D manifold:} Given a set of points in 3D 
digital space,
how would we modify the data set into a manifold without losing or changing the topology (in mathematics)?

\subsection{Boundary Search}

  In general, a point is on the boundary if and only if
it is adjacent to one point in the object and one point not in the object (in 26-connectivity). 
A simple algorithm that goes through each point and tests the neighborhood will determine whether a point is on 
the boundary or not. This is a linear time and $O(log(n))$ space algorithm. 
 
The only thing special is that we use 26-connectivity to determine the boundary points. This is to take all
possible boundary points into consideration in the next step.

\subsection{Determination of the Configuration of Boundary Points} 
 
When all boundary points are found, we need to find their classifications. In other words, we need to determine whether
a special point is in $M_3$, $M_4$, $M_5$, or $M_6$. Here is the problem: if we only have one voxel, is it 
a point (0-cell) or a 3D object (3-cell)? In this paper, we treat it as a 3-cell. 

The input data is in raster space, but the boundary surface will be in point space. We must first make the 
translation. Then, for each point on the surface, we count how many neighbors exist in order to determine its configuration 
category. After that, we use formula (5) to get the genus. 

If we still need to find homology groups, we can just use 
the simple calculations based on Theorem 1,3 to them. Using the program, 
we get the genus$=6$ for a modified real image (Fig. 12).

\begin{figure}[h]
   \begin{center}

   \epsfxsize=3in 
   \epsfbox{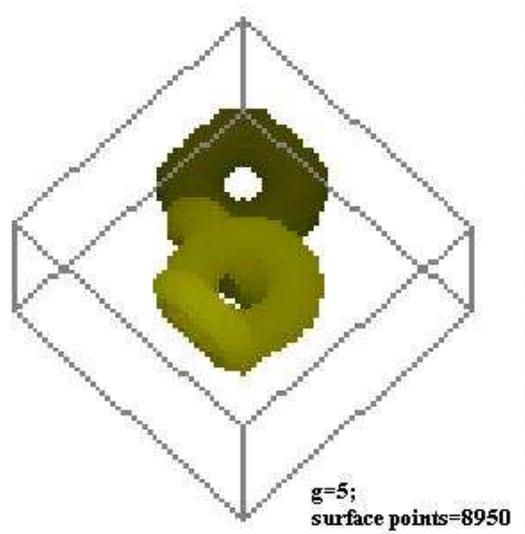}

   \end{center}
\caption{ After Pathological Case Process}
\end{figure}  
 
 \begin{figure}[h]
   \begin{center}

   \epsfxsize=3in 
   \epsfbox{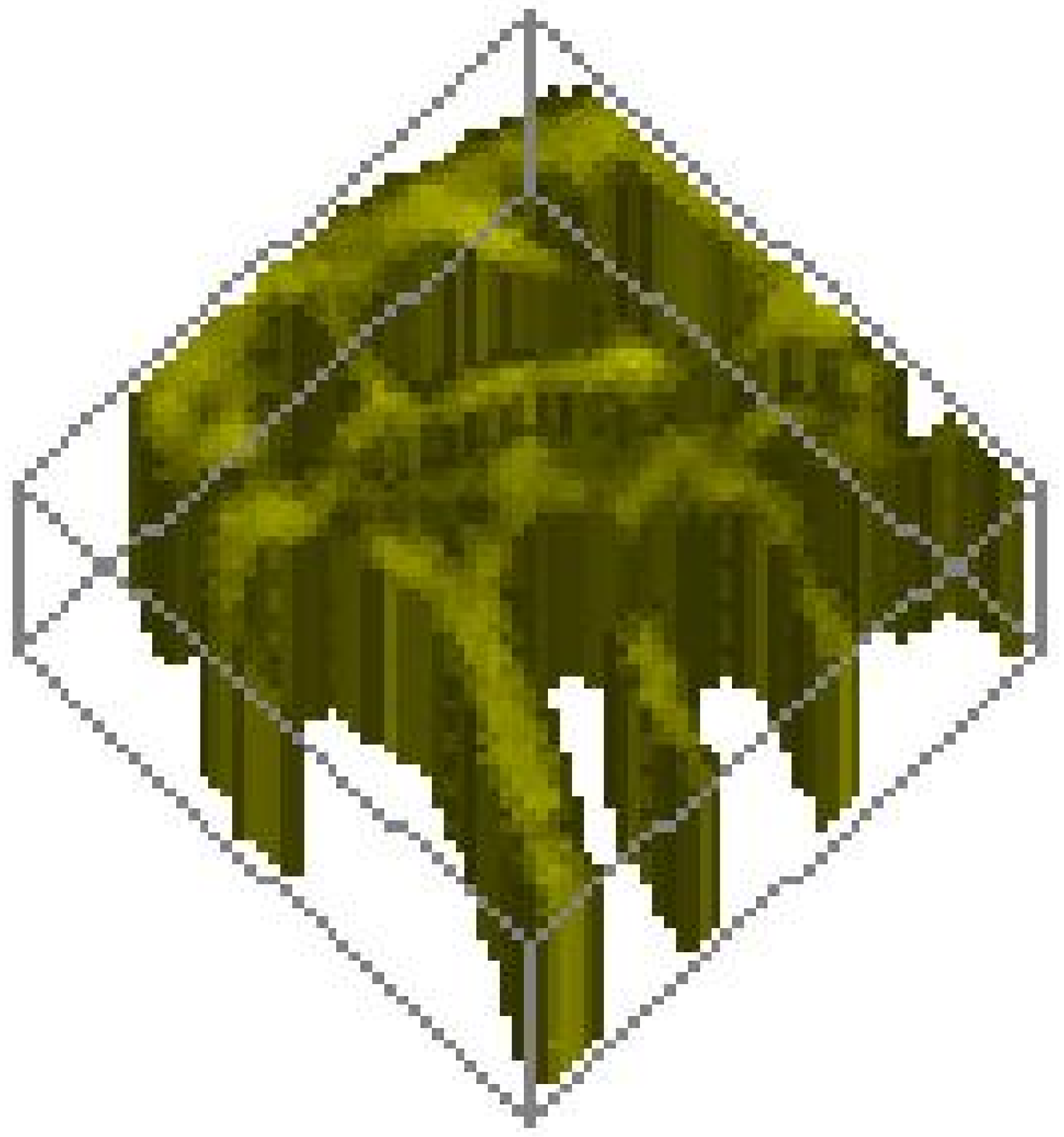}

   \end{center}
\caption{ A Modified Real Image}
\end{figure}

\section{Summary and Discussion} 
In this paper, we have used digital topology to 
get a simple formula for calculating the number of holes
in a connected component in 2D digital space.
The formula is so simple and can be easily implemented.
The author does not know if this formula was known
or obtained already by other researchers.  3D images, we have practically get the genus by extracting
the boundary surfaces and deleting the pathelogical case.
Both algorithms are optimum in terms of time and space
complexity (We usually assume that the number of
 the pathelogical cases are constant). 

This paper is modified based on one unpublished 
note~\cite{Che12} and a conference
paper ~\cite{Che09}.








\end{document}